
\input phyzzx

\overfullrule=0pt

\newif\iffigureexists
\newif\ifepsfloaded
\openin 1 epsf
\ifeof 1 \epsfloadedfalse \else \epsfloadedtrue \fi
\closein 1
\ifepsfloaded
    \input epsf
\else
    \immediate\write20{Warning:
         No EPSF file --- cannot imbed diagrams!!}
\fi
\def\checkex#1 {\relax
    \ifepsfloaded \openin 1 #1
	\ifeof 1 \figureexistsfalse
	\else \figureexiststrue
	\fi \closein 1
    \else \figureexistsfalse
    \fi }
\def\diagram#1 {\vcenter{\def\epsfsize##1##2{0.7##1} \epsfbox{#1}}}
\def\missbox#1{\vbox{\hrule\hbox{\vrule width 0pt height 14pt depth 4pt
	\vrule \kern 4pt #1\kern 4pt \vrule}\hrule }}
\ifepsfloaded
\checkex s7fig3.eps
    \iffigureexists \else
    \immediate\write20{Diagrams are packaged in a separate file}
    \immediate\write20{You should unpack them and TeX again!!}
    \fi
\fi

\def\O{\bf O}
\def\K{\bf K}
\def\Z{\bf Z}

\def\J{{\bf J}}
\def\N{{\bf N}}

\def\l{\lambda}
\def\w{\omega}

\def\fullstop{\,\,.}
\def\komma{\,\,,}
\def\inn{\!\in\!}
\def\is{\!=\!}

\def\pro#1{\!\Buildrel
	\raise 4pt\hbox{\the\scriptscriptfont0 #1}\under\circ\!}
\def\Pro#1{\!\Buildrel#1\under\circ\!}

\def\nicefrac#1#2{\hbox{${#1\over #2}$}}
\def\frac#1/#2{\leavevmode\kern.1em\raise.5ex
		\hbox{\the\scriptfont0
         	#1}\kern-.1em/\kern-.15em
		\lower.25ex\hbox{\the\scriptfont0 #2}}

\def\wickcontract#1#2{
	\vtop{\baselineskip=0pt\lineskip=2pt
		\ialign{##&##\cr
			$#1$&$#2$\cr
			\hfill\vrule height 3pt depth 0pt
			\leaders\vrule height 1pt depth 0pt\hfill&
			\leaders\vrule height 1pt depth 0pt\hfill
			\vrule height 3pt depth 0pt\hfill\cr
		}}}

\date{February 1994}
\pubnum={\vbox{	\hbox{\hfill G\"oteborg-ITP-94-7}
		\hbox{\hfill HUB-IEP-94/1}
		\hbox{\hfill hep-th/9403028}}}

\titlepage

\title{
\break {\fourteenpoint $S^7$ Current Algebras}
	}
\author{
Martin Cederwall
\foot{e-mail tfemc@fy.chalmers.se}
\address{
Institute for Theoretical Physics\break
Chalmers University of Technology and University of G\"oteborg\break
S-412 96 G\"oteborg, Sweden
	}
{\rm and}
Christian R.~Preitschopf
\foot{e-mail preitsch@physik.hu-berlin.de}
\address{
Fachbereich Physik, Humboldt-Universit\"at
zu Berlin, \break D-10099 Berlin, Germany}
	}
\vfill

\abstract
{\vbox{\narrower\narrower{
We present $S^7$-algebras as generalized
Kac-Moody algebras. A number of free-field representations
is found. We construct the octonionic projective spaces
${\O}P^N$.
}}}

\vskip 2cm

\centerline{\sl Talk given by C.P. at the}
\centerline{\sl XXVIIth International
Symposium on the Theory of Elementary Particles}
\centerline{\sl at Wendisch-Rietz, September 7-11, 1993}

\vfill\endpage

\REF\kugoto{T. Kugo and P. Townsend\journal Nucl.Phys.&B221 (83) 357.}
\REF\tentwistor{N.~Berkovits, \sl Phys.Lett. \bf 247B \rm (1990) 45;\nextline
	M.~Cederwall \sl J.Math.Phys. \bf 33 \rm (1992) 388.}
\REF\htwistor{I.~Bengtsson and M.~Cederwall, \sl Nucl.Phys.
	\bf B302 \rm (1988) 81.}
\REF\Berkovitsstring{N. Berkovits, \sl Nucl.Phys. \bf B358 \rm (1991) 169.}
\REF\tentwistor{N.~Berkovits, \sl Phys.Lett. \bf 247B \rm (1990) 45;\nextline
	M.~Cederwall \sl J.Math.Phys. \bf 33 \rm (1992) 388.}
\REF\eightconf{L.~Brink, M.~Cederwall and C.R.~Preitschopf,
	\sl Phys.Lett. \bf 311B \rm (1993) 76.}
\REF\OsipovCentral{E.P.~Osipov, {\sl Lett.Math.Phys.} {\bf 18} (1989) 35.}
\REF\OsipovSugawara{E.P.~Osipov, {\sl Phys.Lett.} {\bf 214B} (1988) 371.}
\REF\sevensphere{M.~Cederwall and C.R.~Preitschopf,
G\"oteborg-ITP-93-34, hep-th/9309030}

\chapter{Introduction}

The seven-sphere is intricately connected to
supersymmetry in high dimensions.
This relation is rather direct via the representation of
Gamma-matrices in terms of octonions [\kugoto].
The algebra of imaginary octonions in turn carries a natural
geometric structure, namely $S^7$.
At the same time we know that in the study of large
supersymmetries exceptional groups occur at various
points and those are also economically described in terms
of octonions. It would be quite satisfying to find a
symmetry principle that explains those coincidences.
A study of the manifestly supersymmetric string
in a twistor-like formulation [\htwistor,\Berkovitsstring,\tentwistor]
as well as in the light-cone [\eightconf]
formulation leads us to believe that such a symmetry
principle could also help a lot in the quantization
of the superstring.
Hence we decided to study in more detail the common
denominator, namely the octonions and their
geometric structure, in the setting of conformal field theory
of $S^7$ and on $S^7$ [\sevensphere].

\chapter{Octonions and $S^7$}

An algebraic way to describe the octonions
consists of writing their multiplication table:
for
$$
x=\sum_{a=0}^7 x_a e_a \in \O \eqn\octo
$$
we set $e_0=1$ and write
$$
x =[x]+\sum_{i=1}^7x_ie_i=[x] + \{x\}		\eqn\decomp
$$
with
$$e_ie_j=-\delta_{ij}+\sigma_{ijk}e_k\komma\eqn\multrule$$
where
the structure constants $\sigma$ are completely antisymmetric
and defined by
$$e_ie_{i+1}=e_{i+3}\ \ .\eqn\eeprod$$
We have set $e_{i+7}=e_i$. One verifies
$$\eqalign{[e_i,e_j]&=2\sigma_{ijk}e_k\cr
	[e_i,e_j,e_k]&=(e_i e_j)e_k - e_i (e_j e_k)
                      =2\rho_{ijkl}e_l\komma\cr
	\rho_{ijkl}&=-(^*\sigma)_{ijkl}=
		-{\nicefrac 1 6}\epsilon_{ijklmnp}\sigma_{mnp}
		\fullstop}\eqn\commass$$
The associator $\rho_{ijkl}$ appears here as a purely
algebraic object. Its geometric meaning becomes clear
if we define the sevenbein $Xe_i$ on the sphere
$S^7\is\{{X\inn\O\ |\ |X|=1}$, where
it defines a globally parallelizing connection.
This means that the connection has zero
curvature, but in our case there is torsion:
$$\half [D_i,D_j]\ = \ T_{ijk}(X)\ D_k \
=\ [(e^*_iX^*)(Xe_j)e_k^*] \ D_k \quad ,\eqn\torsioncomp$$
where $D_i$ is the covariant derivative associated with
the connection we just defined.
Obviously $T_{ijk}(1)= -\sigma_{ijk}$. Less obvious
is the fact that
$$ D_i T_{jkl}(1) \ = \ -2 \rho_{ijkl} \ ,\eqn\covderiv$$
but it is precisely this circumstance that
permits the following construction.

\chapter{Current Algebra}

We define
$$
\eqalign{
\J^i_{(\l)}(z) \ &= \ [\l^* \w e^i] \cr
\J^i_{(S)}(z) \ &= \ \half [(X S)^* (X S) e^i ]\cr
\J^i_{(\l')}(z) \ &= \ [ (X\l')^* (X\w') e^i ]\cr
\J^i_{(b)}(z) \ &= \ [(Xb)^* (Xc) e^i ] \cr
}
\eqn\jcompon
$$
with free bosonic fields
$$
\eqalign{
\wickcontract{\l_a(z) \ }{\w_b(w)} \ &= \  { \delta_{ab} \over z-w }\ ,\cr
\wickcontract{\l_a'(z) \ }{\w_b'(w)} \ &= \  { \delta_{ab} \over z-w }\ ,\cr
}\eqn\freebos
$$
$\l= \l_a e_a$, etc.,  $X= \l/ |\l|$, and
free fermions
$$
\eqalign{
\wickcontract{S_a(z) \ }{ S_b(w)} \ &= \  { \delta_{ab} \over z-w }\ ,\cr
\wickcontract{b_a(z) \ }{c_b(w)}  \ &= \  { \delta_{ab} \over z-w }\ .\cr
}\eqn\freeferm
$$
In fact, we can be a little more general and give the fields
$S_a(z)$, $\l_a'(z), \w_b'(w)$ and $b_a(z), c_b(w)$ an
extra label, say $m$, running from 1 to $N_s$, $N_{\l'}$
and $N_b$, respectively.
Then the current
$$
\eqalign{
\J^i(z) \ = \
\J^i_{\l}(z) \ &+ \
\sum_{m=1}^{N_s} \ \left( \J^i_{(S) m}(z)
\ + \ 2 [X^* \partial X e^i] \right) \cr
\ &+ \
\sum_{m=1}^{N_{\l'}} \ \left( \J^i_{(\l') m}(z)
\ - \ 4 [X^* \partial X e^i] \right) \cr
\ &+ \
\sum_{m=1}^{N_{b}} \ \left( \J^i_{(b) m}(z)
\ + \ 4 [X^* \partial X e^i] \right) \cr
}
\eqn\totalcurr
$$
satisfies the algebra
$$
\wickcontract{\J^i(z) \  }{\J^j(w)} \ =
{8 \delta_{ij} \over (z-w)^2}
\ + \ {2 \over z-w} \ T_{ijk}(X(w))\ \J^k(w)  \ \ ,
\eqn\curralg
$$
where $T_{ijk}(X)$
is the torsion tensor we defined above.
The somewhat unusual form of the current \totalcurr\
arises because we demand \curralg\ without
additional terms built from $X$'s and derivatives, which
generically arise in such a construction.
This condition also fixes the coefficient 8 uniquely.
Our construction can be understood as follows:
a current such as $\J^i_{(S)}$ at $X=1$
would define a Kac-Moody algebra on $S^7$
in exact analogy to the Lie algebra case. We know, however,
that $S^7$ is not a group manifold, and hence
this naive algebra does not close. It misses closing
precisely by terms proportional to the associator
$\rho_{ijkl}$ of the octonions. The first part
$\J^i_{\l}(z)$ of the total current acts as a covariant
derivative on $\J^i_{(S)}$ with respect to the coordinates $X$
of the seven-sphere. The relative coefficients are arranged
such that those two contributions cancel, by
virtue of \covderiv. This mechanism provides us with
the classical version of the current algebra \curralg.
The quantum version arises via a cancellation of
undesirable terms from quantum contributions, namely
double contractions between, say,
$\J^i_{(S)}$ and $\J^j_{(S)}$, and from classical contributions that stem
from a single contraction between $\J^i_{\l}$ and
$[X^* \partial X e^j]$. This type of mechanism is of
course also familiar from the theory of Kac-Moody algebras.

\chapter{Some more complicated Constructions}

Let is now consider, for the moment at a purely classical
level, currents of the type
$$
\J^a(z) \ = \ \sum_{m=0}^{N} \
[\ \w_{(m)}^* \ ( \l_{(m)} X_{(0)}^* ) \ ( X_{(0)} e^{a *} ) \ ] \quad,
\eqn\ncurrent
$$
which generalizes
\totalcurr\ if we set $\l_0 = \l$. Note that
$\J^0(z)$ generates real scale transformations. If we restrict
ourselves to the real, complex or quaternionic cases
and to the zero mode sector of our conformal fields, we may
define the respective projective spaces ${\K} P^N$ by gauging
$\J$, \ie\ by considering the space of homogeneous coordinates
$\l$, ${\K}^{N+1}$, modulo the transformations generated by $\J$. We note that
in those cases all the fields $\l_{(m)}$ are equivalent, since
the $ X_{(0)} $-factors simply drop out.
Octonion multiplication is nonassociative, and hence
the general structure
of such currents can be much more complicated.

If $N=1$, we have two possibilities,
$$
\J^a \ = \ [\ \w_{(0)}^*\ \l_{(0)}\ e^{a *}] \ + \
[\ \w_{(1)}^*\  ( \l_{(1)} X_{(0)}^* )\ ( X_{(0)} e^{a *} ) \ ] \quad
\eqn\currone
$$
or
$$
\eqalign{
\N^a(z) \ &= \ [\ \w_{(1)}^*\ \l_{(1)}\ e^{a *}\ ] \ + \
[\ \w_{(0)}^*\ ( \l_{(0)} X_{(1)}^* )\ ( X_{(1)} e^{a *} )\  ] \cr
\ &= \ [\ ( \J X_{(0)}^*)\  ( X_{(0)} X_{(1)}^* )\  ( X_{(1)} e^{a *} )\ ]
\cr}
\eqn\currtwo
$$
which we represent by diagrams with points corresponding to
the fields $\l_{(m)}$ and links representing the order of
multiplication with respect to the base field indicated by
a black dot:
$$
\checkex s7fig1.eps
\iffigureexists \diagram s7fig1.eps
\else \missbox{Missing Diagrams} \fi
\eqn\twocurrfig
$$

\noindent
\ncurrent\ then corresponds to
the starlike diagram
$$
\checkex s7fig2.eps
\iffigureexists \diagram s7fig2.eps
\else \missbox{Missing Diagrams} \fi
\eqn\starfig
$$

\noindent
We may generalize it to any tree diagram with $N+1$ points
by associating with the path

$$
\checkex s7fig3.eps
\iffigureexists \diagram s7fig3.eps
\else \missbox{Missing Diagrams} \fi
\eqn\pathfig
$$

\noindent
the product
$$
A \Pro{k\rightarrow 0} B \ = \
\bigg( A X_{(k-1)}^*  \
\Big( ( X_{(k-1)} X_{(k-2)}^* ) \cdots
\big( ( X_{(1)} X_{(0)}^* )
( X_{(0)} B ) \big) \cdots \Big) \bigg) \ .
\eqn\pathprod
$$
We note that $|A \Pro{k\rightarrow 0} B| = |A||B|$
and that the product collapses in the real, complex and
quaternionic case to $AB$.
The total classical current that corresponds to a tree diagram
with N+1 points and base point 0 reads now
$$
J^a \ = \ \sum_{m=0}^N \
[ \  \w_{(m)}^*\  \l_{(m)} \Pro {m\rightarrow 0}  e^{a *} \ ] \quad .
\eqn\graphcurr
$$
We may even put fermions at the endpoints of our tree diagram.
Remarkably enough, the current
$$
\eqalign{
J^i \ = \ \ \ \ &[ \  \w_{(0)}^*\  \l_{(0)}\  e^{i *} \ ] \cr
+ \ &\sum_{\{\l\}} \
\left( [ \ \w_{(m)}^*\  \l_{(m)} \Pro {m\rightarrow 0}  e^{i *} \ ]
\ - \ 4 [ \  \partial X_{(m-1)}^*\ X_{(m-1)}
\Pro {(m-1)\rightarrow 0}  e^{i *} \ ] \right)
\cr
+ \ &\sum_{\{S\}} \
\left([ \  S_{m}^*\  S_{m} \Pro {m\rightarrow 0}  e^{i *} \ ]
\ + \ 2\ [ \  \partial X_{(m-1)}^*\ X_{(m-1)}
\Pro {(m-1)\rightarrow 0}  e^{i *} \ ] \right)
\cr
+ \ &\sum_{\{b\}} \
\left([ \  b_{m}^*\  c_{m} \Pro {m\rightarrow 0}  e^{i *} \ ]
\ + \ 4\ [ \  \partial X_{(m-1)}^*\ X_{(m-1)}
\Pro {(m-1)\rightarrow 0}  e^{i *} \ ]  \right)\cr
}
\eqn\gencurr
$$
still generates the algebra \curralg\ at the quantum mechanical level.
Given the complexity of the octonion products in $J^i$, the fact that we
obtain such a simple operator product may safely be regarded as a
miracle.

\chapter{Octonionic Projective Spaces}

At the beginning of the previous section we already
described how we wish to think about projective spaces.
For definiteness we choose our current corresponding to
the starlike diagram \starfig. The gauge orbits of $\J^a$
in ${\O}^{N+1}$ are given by
$$
\Bigg\{ \ \left(\ \l_{(0)} \Pro {X_{(0)}} \ \Omega\ ,\
\cdots \ ,\ \l_{(N)} \Pro {X_{(0)}} \ \Omega \ \right) \ \Bigg|
\ \l_{(m)},\Omega \in {\O}\ , \ |\Omega| =1 \ \Bigg\} \ ,
\eqn\homocoor
$$
where $A \pro X B = (A X^{-1} )( XB)$.
Unfortunately, they are not well defined at $ \l_{(0)} = 0$,
and hence this way of defining ${\O}P^{N}$ is not terribly convenient.
Let us, therefore, fix our gauge freedom and consider a description in
terms of $N+1$ coordinate patches
labelled by the index $b\in \{0,\cdots,N\}$, obtained by
choosing $\Omega$ such that $x^a_a=1$ for
any $a$. This means
$$
x^a_b \ = \ \l_{(a)} \ \Pro {X_{(0)}} \ (\l_{(b)} )^{-1}\ ,
\ a\in \{0,\cdots,N\}
\eqn\homocoor
$$
and implies the transition functions
$$
x^a_b \ = \ 	(x^a_c\ (x^0_c)^{-1} \ )\
		( \ x^0_c\ (x^b_c)^{-1} \ )
	 \ = \ x^a_c\ \pro {(c)} \ (x^b_c)^{-1}
\quad
\eqn\transistarfun
$$
between patch $(b)$ and patch $(c)$.
On each patch we obtain a complete set of independent,
compatible, gauge invariant coordinates.
For a generic diagram we associate with each point $(a)$
one coordinate $x^a_b$ in every patch $(b)$ and
the transition functions read:
$$
x^a_b  \ = \ x^a_c\ \Pro {a \rightarrow b} \ (x^b_c)^{-1}
\quad ,
\eqn\transifun
$$
where the path dependent multiplication is of course performed
in patch $c$. The coordinate patches from different diagrams
are obviously mutually compatible if we identify precisely
one patch of a particular diagram with one patch of every
other diagram. This implies that different diagrams generate
the same manifold.
The diagram \starfig\ is singled out
by the especially simple transition functions involving
patch (0):
$$
x^a_b \ = \ x^a_0\ (x^b_0)^{-1}
\quad ,
\eqn\transone
$$
which are identical to the transition functions in the
associative case.

For ${\O}P^1$ and ${\O}P^2$ we obtain the standard descriptions,
\eg\ the atlas
$$
\eqalign{
	&(1,y_0,z_0)\cr
	&(x_1,1,z_1)\cr
	&(x_2,y_2,1)\cr
}
\eqn\twocharts
$$
with the transition functions
$$
\matrix{&x_1=y_0^{-1}\hfill&&z_1=z_0y_0^{-1}\hfill\cr
	&x_2=x_1z_1^{-1}\hfill&y_2=z_1^{-1}\hfill&\cr
	&&y_0=y_2x_2^{-1}\hfill&z_0=x_2^{-1}\quad .\hfill\cr}
\eqn\overlap
$$

\noindent
We note that these overlap equations are not unique:
we could have substituted the product $A \pro X B = (A X^{-1} )( XB)$ for the
usual octonion product $AB$ in \overlap\ and still obtained a perfectly fine
projective space, which we will call ${\O}P^2 (X)$.
It is also obvious that due to the small dimensionality
of this space the transition functions are considerably simpler
than in the general case \transifun.
Let us now consider the so far unknown space ${\O}P^3$, with
``atlas''
$$
\eqalign{
	&(1,y_0,z_0,v_0)\cr
	&(x_1,1,z_1,v_1)\cr
	&(x_2,y_2,1,v_2)\cr
	&(x_3,y_3,z_3,1)\cr
}
\eqn\threecharts
$$

\noindent
and transition functions \transistarfun. Since these become
ill-defined for $x_a=0$, it behooves us to explain
what we mean with ``atlas'' in this context.
Let us first write down the transition functions in terms of
the coordinates on patch (1):
$$
\eqalign{
(1,y_0,z_0,v_0)\ &=\ (\ 1, x_1^{-1}, z_1 x_1^{-1}, v_1  x_1^{-1}\ )  \cr
(x_2,y_2,1,v_2)\ &= \ (\  x_1 z_1^{-1} ,  z_1^{-1}, 1 ,
				( v_1  x_1^{-1})( x_1 z_1^{-1})\ )   \cr
(x_3,y_3,z_3,1)\ &=\ (\  x_1  v_1^{-1}, v_1^{-1},
				( z_1  x_1^{-1})( x_1 v_1^{-1}) , 1\ )  \cr
}
\eqn\threetransone
$$

\noindent
The ``sphere at infinity'' in chart (0) is described
in charts (1), (2) and (3) as
$$
\eqalign{
(x_1,1,z_1,v_1)\ &=\ (\ 0,1, z_1, v_1 )  \cr
(x_2,y_2,1,v_2)\ &= \ (\ 0, z_1^{-1},1,
		(v_1  \hat x_1^{-1})( \hat x_1 z_1^{-1})\ ) \cr
(x_3,y_3,z_3,1)\ &=\ (\ 0, v_1^{-1},
		(z_1  \hat x_1^{-1})( \hat x_1 v_1^{-1}),1\ ) \quad , \cr
}
\eqn\sphereatinf
$$

\noindent
where $\hat x = x/ |x|$. Evidently the sphere is mapped
to ${\O}P^2 (\hat x_1)$, but strictly speaking $\hat x_1$ is
not well defined at $x_1=0$. For the ``sphere at infinity''
all the charts become singular, chart (0) because some of
the cordinates grow without bound, and charts (1), (2) and (3)
because they become different projections of the image
of this sphere under the transition functions. In other words,
from the data $z_1$, $v_1$ and $v_2$, for example, we can
determine where we are on this image, but not from the data
of patch (1) alone. In the real, complex and quaternionic cases
the factors $\hat x_1$ in \sphereatinf\ vanish
by associativity and we obtain
the famous Hopf maps as the maps from the ``sphere at infinity''
in patch (0) to the submanifold $x_a=0$, \ie\ to ${\K}P^2$.
In the octonionic case this sphere is mapped to a submanifold
of real dimension $23$. The fibering we obtain is
only a ${\Z}_2$-fibration as in the real case, since we can
solve $z_1$, $v_1$ and $v_2=(v_1  \hat x_1^{-1})( \hat x_1 z_1^{-1})$
for $\hat x_1$ up to factor of (-1).
The analog of the equation
${\K}P^3 = {\K}^3 \cup {\K}^2  \cup {\K}^1 \cup  {\K}^0$
for associative ${\K}$ is now

$$
\eqalign{
{\O}P^3 \ &=\ {\O}^3 \cup {\O}P^2(S^7) \cr
{\O}P^2 \ &=  {\O}^2 \cup {\O}^1 \cup  {\O}^0 \quad \cr
}
\eqn\homology
$$

\noindent
with ${\O}P^2(S^7) = \{  {\O}P^2(X)\ |\ X \in S^7 \} $.

The generalization of the above construction to ${\O}P^N$ is
straightforward. One obtains ${\O}P^N(S^7)$ if one replaces
the usual octonion product in \transifun\
with an $X$-dependent one and writes instead of
\transistarfun
$$
x^a_b \ = \ 	(x^a_c\ \pro X \ (x^0_c)^{-1} \ )\ \pro X \
		( \ x^0_c\  \pro X \ (x^b_c)^{-1} \ )
\quad .
\eqn\transseven
$$
This is the image of the ``sphere at infinity''
in ${\O}^{N+1}$ as seen in the $N+1$ coordinate patches
$b=1,2,\cdots,N+1$, each of which provides us only
with a projection of this image.

\chapter{Sugawara Construction}

The similarity of $S^7$ to a group manifold and the results of
Osipov [\OsipovSugawara] for Malcev-Kac-Moody algebras
let us expect that a Sugawara construction
is possible also for $\widehat{S^7}$. This is not difficult to
verify for the simplest current ${\J}^i_{\l}$ defined in \jcompon\
and the associated energy-momentum tensor
$$
\eqalign{
L 	&\ = \ - {1\over 8} : \J_j\ \J_j : \cr
	&\ = \ {7\over 8}\ [\ \l^* \partial \w - \partial\l^* \w\ ]
		\ +\  {1\over 8}\ [\ \l^*\w e_j\  ]\ [\ \l^*\w e_j\  ]\ \ , \cr
}
\eqn\sugl
$$
where we have normal-ordered the currents in the standard way:
$$
 : \J_j(w) \J_k(w) :\ = \lim_{z\rightarrow w} \
	\left( \J_j(z) \J_k(w) - {56\over (z-w)^2} \right)\ .
\eqn\normalone
$$
In order to generalize this result to currents with arbitrary
field content, one has to be careful about normal ordering.
Up to this point we have implicitly assumed free-field
normal ordering on the right-hand side of the current-current
commutation relations. Even though the torsion tensor
$T_{ijk}(X)$ commutes with $\J_k$, the product of those
two operators obeys
$$
\eqalign{
T_{ijk}(X)\J_k &\ =\  : T_{ijk}(X)\ \J_k :
+\ 8\ T_{ijk}(X)\ [X^* \partial X e_k]\cr
&\ =\ : T_{ijk}(X)\ \J_k :\ -\ \ T_{ikl}(X)\ \partial_z T_{jkl}(X) \  ,\cr
}
\eqn\normaltwo
$$
where the left-hand side is free-field normal ordered and
the first term on the right-hand side is current normal ordered.
It is useful to employ the technology described by
Bais, Bouwknegt, Surridge and Schoutens to show the
following result:
for currents that obey
$$
\wickcontract{\J_i(z)\ }{ \J_j(w)}\ =\ -{k\ \delta^{ij} \over (z-w)^2} +
{2\over z-w}  : T_{ijk}(X)
\bigl( \J_k - 8\  [\ X^* \partial X e_k\ ]\bigr )(w):
\ ,
\eqn\currope
$$
the Sugawara energy-momentum tensor is given by
$$
L\ =\ - {1\over 2k+24} : \J_j \J_j :
\eqn\sugj
$$
and has a central charge of
$$
c =- {7k \over k+12} \ ,
\eqn\sugjcent
$$
in accordance with the known results for Kac-Moody algebras.
\sugj\ and \sugjcent\ are equivalent to Osipov's
formulas [\OsipovCentral]. For currents that obey \curralg\ we obtain
unfortunately a negative value of $c$, in accordance
with the fact that the simplest system satisfying
\curralg\ is constructed from a pair of bosonic ghosts.
We note that in contrast to
the Kac-Moody case, the requirement that
$\J_i(z) + \alpha [X^* \partial X e_i](z)$ should transform
like a dimension 1 current under the action of the Sugawara
energy-momentum tensor fixes only the constant $\alpha$,
not the precise form of the stress-energy tensor.
There is a three-parameter family of candidate Sugawara
tensors which satisfy this condition. The operator product
of $L(z)$ with itself determines the parameters. There are
two solutions: \sugj\ for any $k$ and another solution
which would requires a complex value of $k$. The Sugawara
construction is in this sense unique.

\chapter{BRST}

In order to construct the BRST charge $Q$ for the gauge
symmetry generators $\J^i$, we introduce ghost fields
$B^i$ and $C^j$ of dimension 1 and 0 respectively, with
correlator
$$
\wickcontract{B^i(z) \ }{ C^j(w)} \ = \  {\delta^{ij} \over z-w} \ ,
\eqn\ghostcorr
$$
and construct the BRST current in the usual manner:
$$
q(z) \ = \ C^i \N^i \ - \ T^{ijk}(X) C^i C^j B^k \ ,
\eqn\brstcurr
$$
where the currents $\N^i = \J^i + 8 [X^* \partial X e^i]$
have to obey
$$
\wickcontract{\N^i(z) \ }{ \N^j(w)} \ =
{24 \delta_{ij} \over (z-w)^2}
\ + \ {2 \over z-w} \ : T_{ijk}(X(w))\ \N^k(w) : \ \
\eqn\ncurralg
$$
if the BRST charge $Q=\oint {dz \over 2\pi i}\ q(z)$
is to be nilpotent. Note the current normal-ordering
in the above equation, which is satisfied if
the currents $\J^i$ generate \curralg, \ie\ for
all the quantum currents constructed in the previous
sections.

We have to mention one unusual property of the BRST charge $Q$:
the current obtained in the canonical way as the anticommututator
$\{B^i(z), Q\}$ does not generate the algebra \curralg. In fact, we have
not been able to find a representation of \curralg\ with seven
conjugate fields along the lines describe above.

\chapter{Outlook}

In this paper we have shown that $S^7$ provides us with similar
structures as $S^3$. Clearly much remains to be done:
a representation theory of $S^7$ needs to be worked out,
octonionic manifolds should be studied and their relation
to N=8 superconformal algebras should be established.
Octonionic superconformal field theory must be explored.
On these and other related issues we hope to report in
the near future.

\ack
We would like to
thank Nathan Berkovits, Lars Brink
and Peter West for valuable discussions.

\refout

\endpage
\end